%% file: main.tex
\documentclass[letter]{article}

\usepackage{jheppub} % for details on the use of the package, please
\usepackage{graphicx} % Required for inserting images\
\usepackage[utf8]{inputenc}
\usepackage{amssymb, amsmath}
\usepackage{import}
\usepackage{pdfpages}
\usepackage{bm}
\usepackage{mathtools}
\usepackage{bbold}
\usepackage{hyperref}
\usepackage{tensor}
\usepackage{cancel}
\usepackage{braket}
\usepackage{tikz}

\newcommand{\dd}{\mathrm{d}}
\newcommand{\OO}{\mathcal{O}}
\newcommand{\ocft}{\mathcal{O}_{\mathrm{CFT}}}
\newcommand{\ZZ}{\mathbb{Z}}
\newcommand{\subtx}[1]{_{\mathrm{#1}}}

\newlength{\wdth}

\def\beq{\begin{equation}}
\def\eeq#1{\label{#1}\end{equation}}
\def\eeqn{\end{equation}}
\def\beqa{\begin{eqnarray}}
\def\eeqa#1{\label{#1}\end{eqnarray}}
\def\eeqan{\end{eqnarray}}

\def\leqn#1{(\ref{#1})}

\def\lsim{\mathrel{\raise.3ex\hbox{$<$\kern-.75em\lower1ex\hbox{$\sim$}}}}
\def\gsim{\mathrel{\raise.3ex\hbox{$>$\kern-.75em\lower1ex\hbox{$\sim$}}}}

\title{Conformal Freeze-in Dark Matter: 5D Dual and Phase Transition}
\author{Lillian Luo and Maxim Perelstein}
\affiliation{Department of Physics, LEPP, Cornell University, Ithaca, NY 14853, USA}

\abstract{Conformal Freeze-in (COFI) scenario postulates a dark sector described by a conformal field theory (CFT) at energies above the ``gap scale" in the keV$-$MeV range. At the gap scale, the dark CFT undergoes confinement, and one of the resulting bound states is identified as the dark matter candidate. In this paper, we study this model in the context of the AdS/CFT correspondence with a focus on the mechanism of the infrared (IR) breaking of conformal invariance in the dark sector. We construct the holographic dual to the conformal dark sector, given by a Randall-Sundrum-like model in 5D, where the Standard Model (SM) fields and the dark matter candidate are placed on the ultraviolet (UV) and IR branes respectively. The separation between the UV and IR branes is stabilized by a bulk scalar field, naturally generating a hierarchy between the electroweak scale and the gap scale. We find that the parameter space of COFI comprises two distinct branches of CFT's living on the Anti-de-Sitter (AdS) boundary, each corresponding to a different UV boundary condition. The two branches of CFT's result in different radion potentials. The confinement of the CFT is dual to the spontaneous symmetry breaking by the 5D radion potential. We then use this dual 5D setup to study the cosmological confining phase transition in the dark sector. We find the viable parameter space of the theory which allows the phase transition to complete promptly without significant supercooling.}

\begin{document}

\maketitle

\section{Introduction}

Conformal field theories (CFTs) seem to be ubiquitous on the landscape of consistent quantum field theories. For example, non-trivial interacting fixed points of renormalization group flows are described by  CFTs. Even when mass scales are present, any theory with a parametric separation between the infrared (IR) and the ultraviolet (UV) scales can be thought of as approximately conformal at the intermediate scales. 

Robust observational evidence for the existence of dark matter (DM), and the well--known difficulty of incorporating the DM within the Standard Model (SM), strongly motivate a search for theoretical extensions of the SM which contain fields responsible for the DM. A very generic possibility is that the DM field, along with other fields not charged under the SM gauge symmetries, can form a ``dark" (or sequestered) sector. Many model-building attempts along these lines have been made recently. The above-mentioned ubiquity of the CFT's on the theoretical landscape motivates considering a dark sector that possesses a conformal symmetry, or at lease is approximately conformal within a broad window of energy scales. Such an exploration was initiated in Refs.~\cite{Hong_2020,Redi:2021ipn,Chiu:2022bni,Hong_2023,Hong:2024zsn}. 

Obtaining a dark matter candidate from a conformal dark sector requires that the conformal symmetry be broken in the IR, since the energy density in the conformal regime redshifts as radiation and not matter. This breaking can be achieved in a minimal and elegant way by coupling the dark sector to the SM via a ``portal" interaction. The well-understood breaking of conformal symmetry in the SM (due to electroweak symmetry breaking as well as the non-vanishing beta functions) is then translated to the dark sector by the portal. The same portal interaction may be responsible for the DM production in the early universe, leading to a simple and predictive scenario which was dubbed ``Conformal Freeze-In" (COFI). Several versions of this idea, based on different portal interactions, have been considered~\cite{Hong_2020,Chiu:2022bni,Hong_2023,Hong:2024zsn} and shown to provide a phenomenologically viable DM candidate with a mass in the keV$-$MeV range, which will be probed by the upcoming large-scale structure observations. It should be noted that while the conformal symmetry is broken in the IR, the transfer of energy from the SM to the dark sector, which is ultimately responsible for setting up the DM relic density, occurs at temperatures at which the dark sector is still conformal. As a result, conformal dynamics of the dark sector plays a crucial role in determining the DM relic density. This is especially interesting when the dark CFT is strongly interacting, and can contain operators of non-(half-)integer dimension. The theory is sufficiently constrained by the conformal symmetry to enable calculations of inclusive rates of energy transfer from the SM to the dark sector, and the predicted parametric dependence of the relic density on the DM mass and other parameters is distinct from any weakly-coupled DM model.         
AdS/CFT duality provides a powerful theoretical tool for studying strongly-coupled dynamics in (approximately) conformal four-dimensional theories. Duality relates observables in a 4D CFT to those in a weakly-coupled five-dimensional (5D) theory living on an Anti-de-Sitter (AdS) background space. The studies of the AdS/CFT duality were pioneered in the context of $\mathcal{N}=4$ super-Yang-Mills theory and supergravity, but have been subsequently extended to non-supersymmetric, phenomenologically motivated models such as the Randall-Sundrum (RS) model~\cite{Randall_1999,Rattazzi_2001,Arkani-Hamed:2000ijo}. The studies of the Conformal Freeze-In model for dark matter have so far been performed only within the 4D context. The main goal of this paper is to construct a 5D dual of this model. This construction will enable further studies of the scenario in a weakly-coupled 5D framework.  

One aspect of the COFI scenario which has not been studied in detail so far is the cosmological phase transition which occurs when the temperature drops below the IR scale of the dark-sector theory. The dark sector transitions from the high-temperature conformal phase into a low-temperature confined phase populated by the bound states, including the DM particle. As an application of our construction, we will use our 5D dual description to study this phase transition. This study will draw on the extensive literature which considered the phase transition in the RS models~\cite{Creminelli_2002,Randall:2006py,Konstandin:2011dr,vonHarling:2017yew,Baratella:2018pxi,Agashe_2020,Agashe_2021,Bruggisser:2022rdm,Eroncel:2023uqf,Mishra:2023kiu,Mishra:2024ehr}. Notably, the RS phase transition occurs at the TeV scale in the observable sector, while the COFI transition takes place at temperatures of order MeV in the dark sector. However, in most other respects these two cases are very similar. Due to the structure of our 5D dual theory, the COFI phase transition is especially closely related to that in the Relevant Dilaton Stabilization (RDS) model~\cite{Cs_ki_2023}. Similar to the RDS, and unlike the canonical Goldberger-Wise model of stabilization in the RS framework~\cite{GOLDBERGER2000275}, the phase transition in the COFI model will generically complete without significant supercooling, leading to phenomenologically viable cosmology.   

The rest of the paper is organized as follows. In Section~\ref{sec:review} we provide a brief recap of the conformal freeze-in (COFI) scenario of dark matter production, based on the earlier studies in the 4D language. In preparation for constructing the 5D dual, in Section~\ref{sec:CFT} we describe the conformal symmetry breaking in the COFI scenario in the language of spontaneous symmetry breaking, using the spurion approach. Section~\ref{sec:dual} describes the proposed 5D dual of the COFI scenario. (Some of the related technical details are relegated to the Appendix~\ref{sec:app}.) Using this dual description, we study the conformal symmetry-breaking phase transition in Section~\ref{sec:PT}. Finally, we present our conclusions in Section~\ref{sec:conc}.    

\section{Brief Review of the COFI Scenario}
\label{sec:review}

Consider a dark sector described by a CFT below an ultraviolet scale $\Lambda_{\rm UV}$. We will assume that $\Lambda_{\rm UV}$ lies well above the SM electroweak scale. The COFI scenario is agnostic with respect to the nature of the dark sector above $\Lambda_{\rm UV}$; it may, for example, be described by a gauge theory flowing to a strongly-interacting IR fixed point, a la Banks-Zacks. The CFT is assumed to contain a relevant operator ${\cal O}_{\rm CFT}$ of dimension $d<4$. This operator is coupled to the SM sector via a ``portal" interaction:
\begin{equation}
    \mathcal{L}\subtx{int} = \dfrac{\lambda\subtx{CFT}}{\Lambda\subtx{CFT}^{D-4}}\OO\subtx{SM}\ocft\,, \label{eq:dl}
\end{equation}
where $\lambda_{\rm CFT}$ is a small dimensionless constant, ${\cal O}_{\rm SM}$ is a gauge-invariant SM operator of dimension $d_{\rm SM}$, and
$D=d+d_{\rm SM}$. In this paper, we will focus on the Higgs portal~\cite{Hong_2020}:\footnote{For all other Lorentz-scalar renormalizable SM operator portals studied in Ref.~\cite{Hong_2023}, the dominant conformal symmetry-breaking effect in the IR comes from their mixing with the Higgs portal operator due to loop effects. As a result, the physics considered in this paper would be very similar in all those cases.}
\beq
{\cal O}_{\rm SM}=|H|^2,~~~~~~~~d_{\rm SM}=2.
\eeq{HP}
Below the weak scale, the Higgs boson can be integrated out, and the portal interaction generates a CFT deformation of the form
\beq
 \mathcal{L}\subtx{int} \rightarrow \delta {\mathcal L}= \dfrac{\lambda\subtx{CFT}\ v^2}{\Lambda\subtx{CFT}^{D-4}} \ocft\,.
\eeq{HPlow}
This leads to breaking of the CFT at an IR scale 
\beq
M_{\rm gap} \sim \left( \dfrac{\lambda\subtx{CFT}\ v^2}{\Lambda\subtx{CFT}^{d-2}}\right)^{1/(4-d)}\,.
\eeq{Mgap}
At this scale, the dark-sector theory confines, and bound states form. The COFI scenario assumes that one of these bound states is stable, and plays the role of Cold Dark Matter (CDM). 

The COFI scenario assumes that the inflaton decays after the end of inflation reheat the SM but not the dark sector. The dark sector is then populated by scattering and decay of the Higgs bosons, via the portal interaction. After the dark sector undergoes a phase transition at temperature $T_d \sim M_{\rm gap}$, its energy density is transferred to the bound states, and eventually ends up in the form of dark matter particles. The COFI scenario is remarkably predictive: the dark matter relic density, DM particle mass, and other observables are predicted in terms of just two parameters, the portal interaction coefficient $\lambda_{CFT}/\Lambda_{\rm CFT}^{D-4}$ and the CFT operator dimension $d$. It was shown in Ref.~\cite{Hong_2020} that the observed relic density can be reproduced for $1\leq d \lsim 2.5$ and $M_{\rm gap} \sim 1-10$~MeV. Here the lower bound on $d$ is due to unitarity of a scalar CFT operator, while the upper bound is driven by phenomenological considerations. Generically, one expects the mass of the DM candidate to be around $M_{\rm gap}$; however, phenomenological constraints, specifically bounds on DM self-interaction cross section, require a modest hierarchy between these two scales: $M_{\rm dm}/M_{\rm gap} \sim 0.01-0.1$. This hierarchy can be naturally explained if the DM state is a pseudo-Nambu-Goldstone boson (PNGB) of a spontaneously broken global symmetry in the confined phase of the dark sector, similar to the familiar pion in QCD. DM self-interactions, as well as its interactions with the SM, are mediated by a boson with mass around $M_{\rm gap}$, which may be thought of as a dark counterpart of the $\rho$-meson. Ref.~\cite{Hong_2023} performed a phenomenological analysis of two possible scenarios, a scalar (spin-0) or vector (spin-1) mediator.       

\section{CFT Description}
\label{sec:CFT}

As outlined above, the electroweak symmetry breaking in the SM sector induces a small, relevant CFT deformation in the COFI scenario,
\begin{equation}
    \delta\mathcal{L} = c\,\ocft\,.
\end{equation}
The conformal invariance is broken in the IR due to the running of $c$, given as a function of the renormalization scale by 
\begin{equation}
    c(\mu) = c\subtx{UV}(\mu\subtx{UV}/\mu)^{(4-d)}\,, \label{eq:run}
\end{equation}
where $c_\mathrm{UV}\equiv c(\mu_\mathrm{UV})$ is the coupling in the UV. In Higgs portal, 
the value of the coefficient at the weak (UV) scale is 
\beq
c_{UV} = \dfrac{\lambda_{CFT}}{\Lambda_{CFT}^{D-4}}v^2\,.
\eeq{cuv}

A more systematic way to treat conformal symmetry breaking in this setup is provided by spurion analysis. We promote $c$ to an operator with scaling dimension $4-d$ that is also $\ZZ_2$ odd to restore conformal invariance and the $\ZZ_2$ symmetry of the dark sector \cite{Agashe_2020,Cs_ki_2023}. The conformal symmetry is then spontaneously broken by a non-trivial dilaton vacuum expectation value (VEV) set by the effective dilaton potential, 
\begin{equation}
    V\subtx{eff}(\chi) = \lambda \chi^4 - \lambda_2\,\mu\subtx{UV}^{2(4-d)}\chi^{2d-4}\,,\label{eq:V4d}
\end{equation}
where $\lambda$ and $\lambda_2$ are dimensionless constants. In particular, $\lambda_2$ parametrically depends on the small coupling $c$,
\begin{equation}
    \lambda_2 \propto c_\mathrm{UV}^2\mu_\mathrm{UV}^{2(d-4)}\,.
\end{equation}
The potential has a stable minimum at 
\begin{equation}
    \braket{\chi} = \mu_{UV}\left(\dfrac{(d-2) \lambda_{2}}{2\lambda}\right)^{1/2(4-d)} \propto c_\mathrm{UV}^{1/(4-d)}\,,
\end{equation}
when $\lambda_2 > 0$ for $d > 2$, and $\lambda_2 < 0$ for $d < 2$. This is precisely the gap scale $M_\mathrm{gap}$, below which the dark sector confines and form bound states.

We substitute in Eq.~\leqn{cuv} for the Higgs portal and find that the gap scale is
\begin{align}
    M_\mathrm{gap} = \braket{\chi} \sim \left(\frac{\lambda_{CFT}}{\Lambda_{CFT}^{D-4}}v^2\right)^{1/(4-d)}\,,
\end{align}
in agreement with the NDA estimation in Eq.~(\ref{Mgap}) and earlier papers~\cite{Hong_2020,Chiu:2022bni,Hong_2023}.
 
The breaking of conformal invariance may be appropriately described by a dilaton corresponding to spontaneously broken scale invariance, provided that the mass of the dilaton is below the scale of spontaneous breaking, $\braket{\chi}$. Thus, $\lambda$ must satisfy the constraint
\begin{equation}
    8\lambda (4-d) \lesssim 1\,. \label{eq:lamspon}
\end{equation}
This constraint will re-appear below as a consistency condition on the parameters of the 5D dual of the COFI setup.

\section{AdS Dual in 5D}\label{sec:dual}

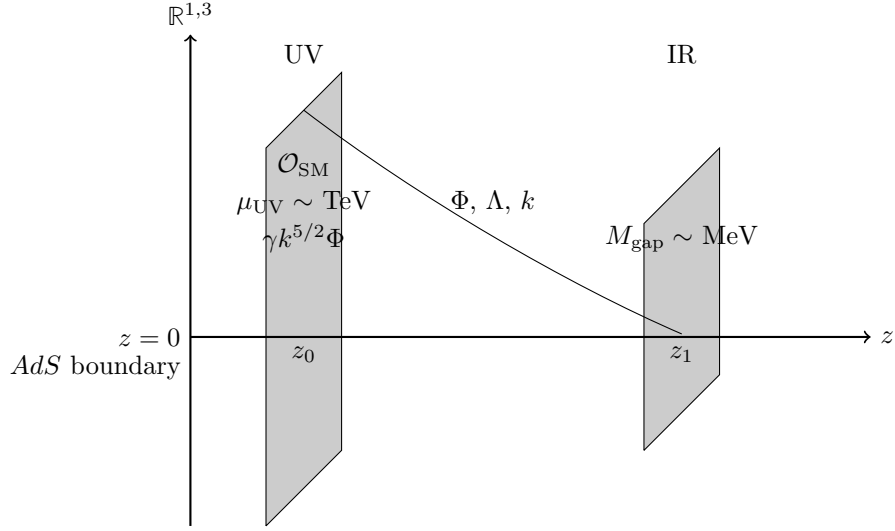
\begin{figure}
    \centering
    \begin{tikzpicture}[draw=black]
        \filldraw[fill=gray!40!white] (0,0) -- (0,5) -- (1,6) -- (1,1) -- (0,0);
        \draw (0.5,6) node[anchor=south,black] {UV};
        \draw (0.5,4.5) node[anchor=south,black] {$\mathcal{O}_\mathrm{SM}$};
        \draw (0.5,4) node[anchor=south,black] {$\mu_\mathrm{UV} \sim$ TeV};
        \draw (0.5,3.5) node[anchor=south,black] {$\gamma k^{5/2} \Phi$};
        \filldraw[fill=gray!40!white] (5,1) -- (5,4) -- (6,5) -- (6,2) -- (5,1);
        \draw (5.5,6) node[anchor=south,black] {IR};
        \draw (5.5,3.5) node[anchor=south,black] {$M_\mathrm{gap} \sim$ MeV};
        \draw (0.5,5.5) .. controls (2,4.35) and (4,3.186) .. (5.5,2.54);
        \draw[thick,->] (-1,2.5) -- (0.5,2.5) node[anchor=north,black] {$z_0$} -- (5.5,2.5) node[anchor=north,black] {$z_1$} -- (8,2.5) node[anchor=west,black] {$z$};
        \draw[thick,->] (-1,0) -- (-1,2.5) node[anchor=east,black] {$z=0$} -- (-1,6.5) node[anchor=south,black] {$\mathbb{R}^{1,3}$};
        \draw (3,4) node[anchor=south,black] {$\Phi$, $\Lambda$, $k$};
        \draw (-1,2.1) node[anchor=east,black] {$AdS$ boundary};
    \end{tikzpicture}
    \caption{The holographic dual of COFI DM realized in an RS-like geometry. $\ocft$ is dual to a bulk scalar field $\Phi$. The small explicit CFT breaking due to the coupling of $\ocft$ to $\OO_\mathrm{SM}$ corresponds to UV localized terms, and the CFT bound states are located on the IR brane.} 
    \label{fig:5D}
\end{figure}

By the AdS/CFT correspondence, the strongly coupled CFT of the dark sector has an equivalent description in a weakly coupled 5D holographic dual qualitatively similar to the Randall-Sundrum model \cite{Randall_1999,Rattazzi_2001,Arkani-Hamed:2000ijo}, depicted in Fig.~\ref{fig:5D}. The CFT operator $\ocft$ is dual to a bulk scalar field $\Phi$, whose bulk mass $m$ is related to the dimension $d$ of $\ocft$ by $d(d-4)=m^2$. The Standard Model fields are localized on the UV brane. The Higgs portal interaction couples a UV-brane localized $|H|^2$ operator to the bulk scalar field $\Phi$. The breaking of scale invariance induced by the deformation $\delta \mathcal{L}$ is realized in 5D when the Higgs field acquires a VEV, triggering a small, UV localized tadpole that sources the field $\Phi$. The IR confining dynamics of the dark sector is modeled by introducing an IR brane. CFT bound states, including the dark matter particle, correspond to IR-brane localized fields. A coupling between $\Phi$ and the IR brane ties the location of the IR brane (and hence the confinement scale) to the UV tadpole. This is similar to the recently proposed ``Relevant Dilaton Stabilization" model~\cite{Cs_ki_2023}, except our construction is at a different energy scale: our UV brane is at a TeV scale (vs. Planck scale in~\cite{Cs_ki_2023}) while our IR brane is at an MeV scale (vs. TeV scale in~\cite{Cs_ki_2023}). Unlike the Goldberger-Wise mechanism \cite{Goldberger_1999}, the stabilization scheme typically adopted in RS models, the UV and IR hierarchy is generated by the small UV tadpole rather than a small bulk mass.

We now explicitly construct the 5D dual of the conformal dark sector. Consider a slice of AdS$_5$ where the extra dimension is compactified by an orbifold symmetry. The UV and IR branes are positioned at the orbifold fixed points at $z=z_0$ and $z=z_1$ respectively. The brane locations set the UV and IR scales of the theory at $\mu\subtx{UV} \equiv 1/z_0 \sim \text{TeV}$ and $\mu\subtx{IR} \equiv 1/z_1 \sim M\subtx{gap}$. The 5D gravitational action is given by
\begin{equation}
    S = -\int \dd^4 x\ \dd z \sqrt{|g|} \bigl(2M_5^3R + \Lambda\bigr) + \sqrt{|g_{\mathrm{ind}}|}\Lambda\subtx{UV}\delta(z-z_0) + \sqrt{|g_{\mathrm{ind}}|}\Lambda\subtx{IR}\delta(z-z_1)\,. \label{eq:EH}
\end{equation}
Assuming small back-reaction from the matter content, the gravitational action yields the Randall-Sundrum solution with metric
\begin{equation}
    \dd s^2 = \frac{1}{(kz)^2}(\tensor{\eta}{_\mu_\nu}\dd x^\mu \dd x^\nu - \dd z^2)\,,
\end{equation}
where $k$ is the AdS curvature. The bulk cosmological constant (CC) $\Lambda = -24M_5^3k^2$ and the brane tensions $\Lambda\subtx{UV}=-\Lambda\subtx{IR}=24M_5^3k$.\\
The holographic dual of $\ocft$ is a bulk scalar field $\Phi$ whose action is given by
\begin{equation}
    S_{\Phi} = \int \dd^4x\ \dd y \sqrt{|g|}\bigl(\frac{1}{2}\tensor{g}{^M^N}\partial_M \Phi \partial_N \Phi - \frac{1}{2}m^2 \Phi^2\bigr) - \sqrt{|g\subtx{ind}|} V\subtx{UV}(\Phi)\delta(z-z_0) - \sqrt{|g\subtx{ind}|}V\subtx{IR}(\Phi)\delta(z-z_1)\,, \label{eq:Sscalar}
\end{equation}
with brane localized potentials 
\begin{equation}
    V\subtx{UV}(\Phi) = \frac{1}{2}m\subtx{UV}\Phi^2 + \gamma k^{5/2}\Phi \text{\quad and \quad} V\subtx{IR}(\Phi) = \frac{1}{2}m\subtx{IR}\Phi^2\,,
\end{equation}
where $\gamma \propto c(\mu_\mathrm{UV}) \mu_\mathrm{UV}^{d-4}$ is a small dimensionless constant set by CFT parameters. We note that the tadpole term, $\gamma k^{5/2} \Phi$, creates an explicit breaking of $\ZZ_2$ symmetry that is otherwise preserved by $\Phi$, allowing for a naturally small $\gamma$ akin to the suppression of the CFT deformation $\delta \mathcal{L}$ in the 4D description. This stabilizes the hierarchy between the UV and IR scales by generating a non-trivial potential for the radion (dilaton) field. In the original proposal of Relevant Dilaton Stabilization \cite{Cs_ki_2023}, the discussion in 5D is restricted to where the scaling dimension of $\ocft$ is $2 < d < 4$. In the following section we will extend the duality to the full range of scaling dimensions $1 < d < 4$, cut off by the unitarity bound. This completes the 5D description of the COFI scenario for all choices of $d$ permitted by phenomenological bounds.

\subsection{Two Boundary Theories}

As established by the AdS/CFT correspondence, (in $k\equiv1$ units) the bulk scalar mass $m$ in AdS is related to the scaling dimension $d$ of its dual CFT operator by
$$d(d-4)=m^2\,.$$
There are two solutions,
$$d_\pm=2 \pm \sqrt{4+m^2} \equiv 2\pm \nu\,,$$
corresponding to two boundary CFT's \cite{Witten:1998qj, Klebanov_1999}. The Breitenlohner-Freedman bound $m^2 \geq -4$ % cite
ensures $d_\pm$ are real. For the $d_-$ root, $m^2 < -3$ ($\nu < 1$) is additionally required to satisfy the unitarity bound $d > 1$. Then in the range $-4 \leq m^2 < -3$ ($0 < \nu \leq 1$), both boundary theories could be obtained from the scalar field solution near the AdS boundary,
\begin{equation}
    \Phi(x, z) = c_+(x) z^{d_+}+c_-(x) z^{d_-}
\end{equation}
by choosing to impose the boundary condition of either $c_+(x) = 0$ or $c_-(x) = 0$, and turning it on as the source for a boundary CFT operator $\OO$. 

In the COFI scenario, the scalar potential on the UV brane, dual to the deformations to the CFT in 4D, fixes the boundary conditions to 
\begin{equation}
    2z_0 \Phi'(x,z_0) = \frac{\partial V_\mathrm{UV}}{\partial \Phi}\Big|_{\Phi(z_0)}\,.
\end{equation} 
We must then consider its modification to the scaling behavior by computing the two-point function $\langle \OO \OO \rangle$ from the on-shell 5D action in the presence of a boundary source $\Phi_b(x)$ \cite{Minces_2000, Hartman_2008, Kaplan_2009}, using the AdS/CFT dictionary,
\begin{equation}
    Z\subtx{AdS}[\Phi_0]=\int_{\Phi_0}\, \mathcal{D}\Phi e^{-S\subtx{bulk}[\Phi]}=\langle e^{\int \dd^4 x\, \Phi_0 \OO} \rangle = Z\subtx{CFT}[\Phi_0]\,,
\end{equation}
in the limit where the IR brane $z\subtx{IR} \rightarrow \infty$.

The metric in the Euclidean Poincare coordinates is 
\begin{equation}
    \dd s^2 = \frac{1}{z^2}(\tensor{\eta}{_\mu_\nu}\dd x^\mu \dd x^\nu + \dd z^2)\,.
\end{equation}

Using the equation of motion and the boundary condition, 
\begin{align} 
    (\Box - m^2)\Phi &= 0\,, \label{eq:eom_euc}\\
    2 z_0 \Phi'(x, z_0)-m\subtx{UV}\Phi(x, z_0) &= \gamma k^{5/2} + \Phi_b(x)\,,
\end{align}
we find the on-shell action,
\begin{align}
    S\subtx{on-shell} &= \frac{1}{2}\int \dd^4 x \dd z \sqrt{g} \left((\partial \Phi)^2 + m^2\Phi^2\right) - \int \dd^4 x \sqrt{g\subtx{ind}}\left(\frac{1}{2}m\subtx{UV}\Phi^2 + \gamma k^{5/2}\Phi\right) \nonumber\\
    &=\frac{1}{2}\int \dd^4 x \dd z \sqrt{g}\left(\nabla_M(\Phi \partial^M \Phi)-\Phi (\Box-m^2)\Phi\right)-S\subtx{bdry} \nonumber\\
    &=\frac{1}{2}\int \dd^4 x\ \sqrt{g\subtx{ind}}(\Phi_b(x)-\gamma k^{5/2})\Phi(x, z_0)\label{eq:onshell}\,.
\end{align}

Under 4D Fourier transform $x \rightarrow p$,
\begin{equation}
    \Phi(x,z) = \int \frac{\dd^4 p}{(2\pi)^2} e^{ipx}\Phi(p,z)\,,
\end{equation}
the equation of motion (\ref{eq:eom_euc}) becomes
\begin{equation}
    (-p^2z^2 + z^2\partial_z^2 - 3z\partial_z-m^2)\Phi(p,z)=0\,,
\end{equation}
which has IR regular solutions of the form
\begin{equation}
    \Phi(p, z) = A(p) z^2 K_\nu(pz)\,,
\end{equation}
where $K_\nu(z)$ is the modified Bessel function. Applying the boundary condition, noting that the tadpole term from the brane potential only contribute to the $p=0$ Fourier mode, we find
\begin{equation}
    \Phi(p, z) = \left(\frac{z^2 K_\nu(pz)}{(2d_- -m\subtx{UV})z_0^2 K_\nu(pz_0)-2z_0^3pK_{\nu-1}(pz_0)}\right)(\Phi_b(p)+\gamma k^{5/2} \delta^4(p)) \label{eq:p_sol}\,.
\end{equation}

We then use the series expansion for $K_\nu$,
\begin{equation}
    K_\nu(z)=\frac{1}{2}\Gamma(\nu)\Gamma(1-\nu)(\frac{z}{2})^{-\nu}\left[\sum_{n\geq0}\frac{(z/2)^{2n}}{n!\Gamma(n+1-\nu)}-(\frac{z}{2})^{2\nu}\sum_{n\geq0}\frac{(z/2)^{2n}}{n!\Gamma(n+1+\nu)}\right]\,,
\end{equation}
and find that to leading order, 
\begin{align} %saddle point approx?
    \langle\OO(p)\OO(-p)\rangle &= -\frac{\delta^2 S\subtx{on-shell}[\Phi_b]}{\delta \Phi_b(p)\delta \Phi_b(-p)} \bigg\rvert_{\Phi_b=0}\\
    &\propto \left[ (2d_- - m\subtx{UV}) + (pz_0)^{2\nu}(2\nu)2^{-2\nu}\Gamma(-\nu)/\Gamma(\nu)\right]^{-1}\,.
\end{align}

\begin{figure}[t]
    \centering
    \includegraphics[width=0.7\linewidth]{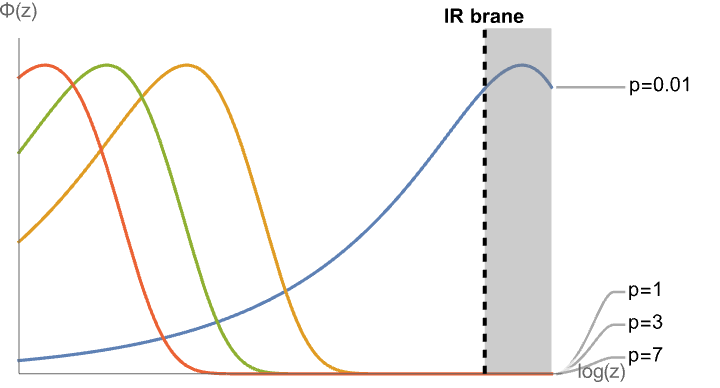}
    \caption{The profile of the scalar Fourier modes along the 5th dimension. The low energy modes ($p \ll 1$) are localized in the IR. In the COFI scenario, the IR brane truncates the 5th dimension, indicated by the dashed line.}
    \label{fig:zprofile}
\end{figure}

Note that from (\ref{eq:onshell}) and (\ref{eq:p_sol}), the on-shell action is quadratic in the boundary source $\Phi_b$, with the tadpole term $\gamma k^{5/2} \Phi$ in the brane localized potential only appearing in lower-order terms. Hence, the scaling behavior of the dual CFT operator is unaffected by the tadpole. The scaling dimension of $\OO$ is thus determined entirely from the UV brane mass term, dual to a deformation of the form $\frac{f}{2}\OO^2$ in the CFT description of the theory, known as a "double-trace deformation". We examine the running of the scaling dimension of the UV boundary CFT to the IR by considering the IR localized low energy modes with $p \ll z_0^{-1}$, visualized in Fig. \ref{fig:zprofile}. For generic values of $m\subtx{UV}$, $p z_0 \ll (2d_- -m\subtx{UV})$, and thus
\begin{equation}
    \langle\OO(p)\OO(-p)\rangle \propto p^{2\nu}\,.
\end{equation}
This means that the scaling dimension of $\OO$ flows to $d_+$ in the IR. When we fine-tune $m\subtx{UV} = 2d_-$, instead the low energy correlation functions
\begin{equation}
    \langle\OO(p)\OO(-p)\rangle \propto p^{-2\nu}\,,
\end{equation}
and so the scaling dimension of $\OO$ remains $d_-$ in the IR\footnote{This is also discussed in related work \cite{Hubisz:2024xnj}.}. In the dual 4D CFT, it is equivalent to tuning the double-trace deformation to $0$.

In COFI, the inclusion of a dynamically generated IR brane from the radion potential, or equivalently, a mass gap from broken conformal invariance, serves as an IR cutoff, so that the scalar modes with $p < \mu\subtx{IR}$ are unphysical, shown in Fig. \ref{fig:zprofile}. Then, it is sufficient to ensure that all Fourier modes with momenta above the IR cutoff have $d_-$ scaling, relaxing the fine-tuning condition in the $d_-$ branch. In summary, constructing the holographic dual to the COFI scenario requires a different choice of the UV-brane localized mass parameter depending on the dimension of $\ocft$. 
For $d>2$ we can choose any $m_\mathrm{UV}$. For $1<d<2$, we must tune $m_\mathrm{UV} \rightarrow 2d_-$.

\subsection{Radion Stabilization}

The stabilizing potential for the radion can be determined from the zero mode solution to the bulk scalar EOM,
\begin{equation}
    \Phi(z)=\Phi_0 (kz)^{2+\nu} + \Phi_1 (kz)^{2-\nu}\,,
\end{equation}
subject to the boundary conditions,
\begin{align}
    2k z_0\Phi'(z_0) &=m\subtx{UV}\Phi(z_0) + \gamma k^{5/2}\,, \nonumber\\
    -2k z_1\Phi'(z_1) &=m\subtx{IR}\Phi(z_1)\,.\label{eq:bvp}
\end{align}
Integrating out the extra dimension from the on-shell action of eq. (\ref{eq:Sscalar}) and substituting in the zero mode solution, we find the effective action
\begin{equation}
    S\subtx{on-shell} = -\int_{z=z_0} \dd^4 x \sqrt{g\subtx{ind}}\frac{1}{2}\gamma k^{5/2}\Phi(z_0) = \int \dd^4 x\frac{(kz_0)^{-4}\gamma^2 k^4 (\tau\subtx{IR}(kz_0)^{2\nu}-(\tau\subtx{IR}+4\nu)(kz_1)^{2\nu})}{2(\tau\subtx{IR}(\tau\subtx{UV}-4\nu)(kz_0)^{2\nu}-\tau\subtx{UV}(\tau\subtx{IR}+4\nu)(kz_1)^{2\nu})}\,,
\end{equation}
where we define dimensionless UV and IR mass mistuning parameters
\begin{equation}
    \tau\subtx{UV} \equiv m\subtx{UV}/k - 2d_-\,, \quad \tau\subtx{IR} \equiv m\subtx{IR}/k + 2d_-\,.
\end{equation}
The radion field is identified as $\chi \equiv \dfrac{1}{z_1}$. Novel to the $d_-$ branch, the fine-tuning condition on the UV mass, $\tau_\mathrm{UV} \ll 1$, leads to a different form of the effective potential than that studied in \cite{Cs_ki_2023}. We now discuss the $d_+$ and $d_-$ cases separately.

\begin{itemize}
    \item $[\ocft] > 2$.\\
    In this sector of the parameter space, we assume $\tau_\mathrm{IR}, \tau_\mathrm{UV} \sim \mathcal{O}(1)$. The effective potential generated by the UV tadpole is
    \begin{align}
        V_+(\chi) &\approx \frac{\gamma^2 k^{4} \tau\subtx{IR}(kz_0)^{2\nu}}{2(\tau\subtx{IR}+4\nu)\tau\subtx{UV}}(kz_1)^{-2\nu}(kz_0)^{-4} + \mathrm{const.}\nonumber\\
        &=\frac{\gamma^2 \tau\subtx{IR}}{2\tau\subtx{UV}(\tau\subtx{IR}+4\nu)}(\frac{1}{z_0})^{4-2\nu} \chi^{2\nu} + \mathrm{const.}\,.
    \end{align}
    We include a $\chi^4$ term generated from the mistuning of the UV and IR brane tension. The effective potential for the radion is then given by
    \begin{equation}
        V\subtx{eff}(\chi) = \frac{24M_5^3}{k^3}\left(\lambda\chi^4 - \lambda_+ \mu\subtx{UV}^{4 - 2\nu} \chi^{2\nu}\right)\,,
    \end{equation}
    where we define the constant coefficient
    \begin{equation}
        \lambda_+ = -\frac{k^3}{24M_5^3}\frac{\tau\subtx{IR}\gamma^2}{2\tau\subtx{UV}(\tau\subtx{IR}+4\nu)}\,.
    \end{equation}

    To ensure the stability of the effective potential, we verify that the scalar zero mode is stable under a small perturbation, $\Phi(z)+\phi(x, z)$. That is, the effective 4D mass of the perturbation $\phi(x, z)$ is positive (non-tachyonic). The mass is found to be $m_\phi^2 \propto (\tau_\mathrm{UV} + \tau_\mathrm{IR}(\mu_\mathrm{UV}z_1)^{-2\nu})$, so we require the UV mistuning parameter $\tau_\mathrm{UV}>0$ to ensure the stability of the zero mode solution.\\

    \item $1 < [\ocft] < 2$.\\
    Much of the derivation remains the same, but with the fine-tuning condition $\tau\subtx{UV} \ll 1$. The potential due to the tadpole is
    \begin{align}
        V_-(\chi) &\approx -\frac{\gamma^2 k^4 (\tau\subtx{IR}+4\nu)}{2\tau\subtx{IR}(4\nu)(kz_0)^{2\nu}}(kz_1)^{2\nu}(kz_0)^{-4} + \mathrm{const.}\nonumber\\
        &=-\frac{\gamma^2 (\tau\subtx{IR}+4\nu)}{2\tau\subtx{IR}(4\nu)}(\frac{1}{z_0})^{4+2\nu} \chi^{-2\nu} + \mathrm{const.}\,.
    \end{align}
    Adding the $\chi^4$ term and defining the constant
    \begin{equation}
        \lambda_- = \frac{k^3}{24M_5^3}\frac{(\tau\subtx{IR}+4\nu)\gamma^2}{2\tau\subtx{IR}(4\nu)}\,,
    \end{equation}
    the effective potential is then
    \begin{equation}
        V\subtx{eff}(\chi) = \frac{24M_5^3}{k^3}\left(\lambda\chi^4 - \lambda_- \mu\subtx{UV}^{4 + 2\nu} \chi^{-2\nu}\right)\,.
    \end{equation}
    The small perturbation of the zero mode has mass $m_\phi^2 \propto ((\tau_\mathrm{IR} + 4\nu) (\mu_\mathrm{UV}z_1)^{2\nu}-4\nu)$, so the effective potential is stable given $(\tau_\mathrm{IR} + 4\nu) > 0$.
\end{itemize}

For all $d$, the effective potential obtained from the 5D construction is consistent with Eq. (\ref{eq:V4d}) obtained from the purely 4D analysis, with $d \equiv 2 \pm \nu$ for the two branches of scaling dimensions. A non-trivial dilaton VEV exists for $\lambda_+ > 0$ and $\lambda_- < 0$ in the two branches of CFT respectively. This further constrains the mistuning parameters: for both branches, $-4\nu < \tau_\mathrm{IR} < 0$. The VEV is given by
\begin{equation}
    \braket{\chi} = \mu_\mathrm{UV}\left(\pm\frac{\nu \lambda_\pm}{2\lambda}\right)^{1/(4\mp 2\nu)} \sim \mu_\mathrm{UV}\gamma^{1/(2\mp\nu)} \sim c(\mu_\mathrm{UV})^{1/(4-d)}\,.
\end{equation}
The small IR scale $\braket{\chi} = M_\mathrm{gap}$ is naturally generated through the small $\gamma$ parameterizing the soft explicit symmetry breaking, which is a feature of the RDS. We can rewrite the potential in terms of the VEV and set $V(\braket{\chi})=0$ (the CC problem is beyond the scope of this paper),
\begin{equation}
    V(\chi) = \frac{24M_5^3}{k^3}\lambda\braket{\chi}^4\left[\frac{\chi^4}{\braket{\chi}^4}-1 \mp \frac{2}{\nu}\left(\left(\frac{\chi}{\braket{\chi}}\right)^{\pm 2\nu} - 1\right)\right]\,. \label{eq:Veff}
\end{equation}

The kinetic term for the dilaton is obtained from the 5D gravitational action by adding a scalar fluctuation to the RS-like background and integrating out the compact dimension, shown in Appendix \ref{ap:kin}. The effective action for the dilaton is then
\begin{equation}
    S_\mathrm{eff} = \int \dd^4 x \left[\frac{3N^2}{4\pi^2}(\partial \chi)^2 - V(\chi)\right]\,,
\end{equation}
where $N$ is the number of degrees of freedom (d.o.f) in the CFT related to the 5D parameters by $N^2 = 16\pi^2(M_5/k)^3$. From the canonically normalized action, the dilaton mass is given by
\begin{equation}
    m_\chi^2 = 8\lambda(4-d)\braket{\chi}^2\,.
\end{equation}
For the phenomenologically viable parameters of the COFI scenario ($d \lesssim 2.4$), the dilaton mass is not suppressed from the gap scale. Thus, additional constraints must be imposed to ensure that the dilaton is the lightest excitation so that the effective theory description is valid. In particular, we require $m_\chi \lesssim m_\mathrm{KK}$, the mass of the Kaluza-Klein modes, which can then be integrated out in the low energy theory. We will again require $m_\chi^2/\braket{\chi}^2 \lesssim 1$ as with Eq. (\ref{eq:lamspon}) so that the scale invariance is mostly broken spontaneously by the IR brane. These consistency constraints lead to a mild tuning on $\lambda \sim 0.01 - 0.1$.

The effective theory analysis is built on the assumption that the backreaction on the gravitational sector is small. We thus also verify that the stress-energy tensor does not receive significant correction from the bulk scalar, and is dominated by the bulk CC and brane tensions \cite{Cs_ki_2023}. Indeed, the contribution from the bulk scalar is proportional to $\lambda M_5^3k$ on the brane and $\lambda M_5^3k^2$ in the bulk, so the mild tuning of $\lambda$ also ensures that the backreaction is negligible.

\begin{figure}
    \centering
    \includegraphics[width=0.5\linewidth]{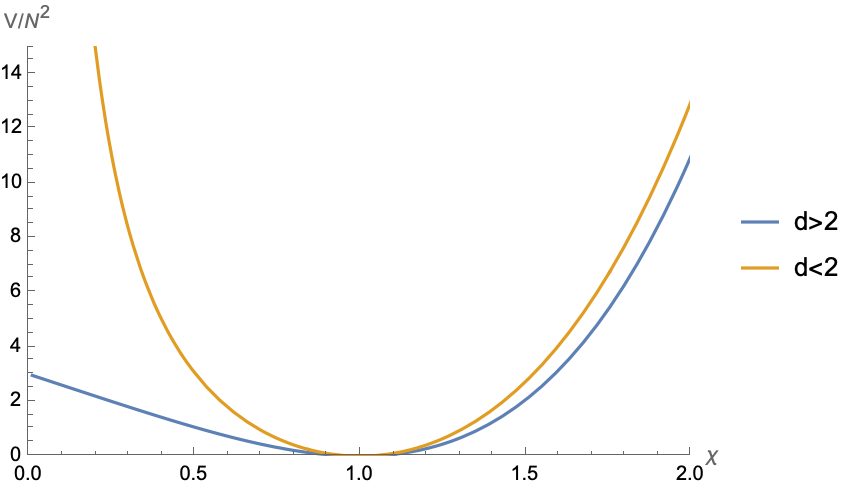}
    \caption{Comparison of the effective potentials for $\ocft$ with $d>2$ (blue) and $1<d<2$ (orange). The potentials are plotted in units $\braket{\chi}=1$. The potential of the $d_-$ branch goes out of the EFT control rapidly as one approaches $\chi=0$.}
    \label{fig:potential}
\end{figure}

The effective potentials for the two branches of boundary CFT are qualitatively different, shown in Fig. \ref{fig:potential}. The potential of the $d_+$ branch resembles that of the Goldberger-Wise mechanism but is steeper \cite{Cs_ki_2023}. On the other hand, the potential of $d_-$ grows rapidly away from the vacuum at $\chi = \braket{\chi}$ and is singular at $\chi = 0$, suggesting that the dilaton effective field theory (EFT) quickly breaks down away from the vacuum.

\subsection{Three-Brane Extension}

\begin{figure}
    \centering
    \begin{tikzpicture}[draw=black]
        \draw[thick,->] (0,0) node[anchor=east,black] {$y=0$} -- (8,0) node[anchor=west,black] {$y$};
        \draw[thick,->] (0,-3) -- (0,3) node[anchor=south,black] {$\mathbb{R}^{1,3}$};
        \draw (0,3.5) node[anchor=south,black] {\textbf{Planck}};
        \draw[thick,<-,blue] (0.1,1) -- (2.2,1);
        \draw[thick,->,blue] (2.7,1) -- (4.9,1);
        \draw (2.5,1) node[anchor=south,black] {$\Lambda_1$};
        \draw (2.5,1) node[anchor=north,black] {$k_1$};
        \draw[thick] (5,-1.5) -- (5,0) node[anchor=north west,black] {$y_I$} -- (5,1.5) node[anchor=south,black] {\textbf{TeV}};
        \draw[thick,<-,blue] (5.1,1) -- (5.5,1);
        \draw[thick,->,blue] (6,1) -- (6.4,1);
        \draw (5.75,1) node[anchor=south,black] {$\Lambda_2$};
        \draw (5.75,1) node[anchor=north,black] {$k_2$};
        \draw[thick] (6.5,-1.3) -- (6.5,0) node[anchor=north west,black] {$y_r$} -- (6.5,1.3) node[anchor=south,black] {\textbf{Dark}};
    \end{tikzpicture}
    \caption{Three-brane extension of the COFI dark sector.}
    \label{fig:UV}
\end{figure}
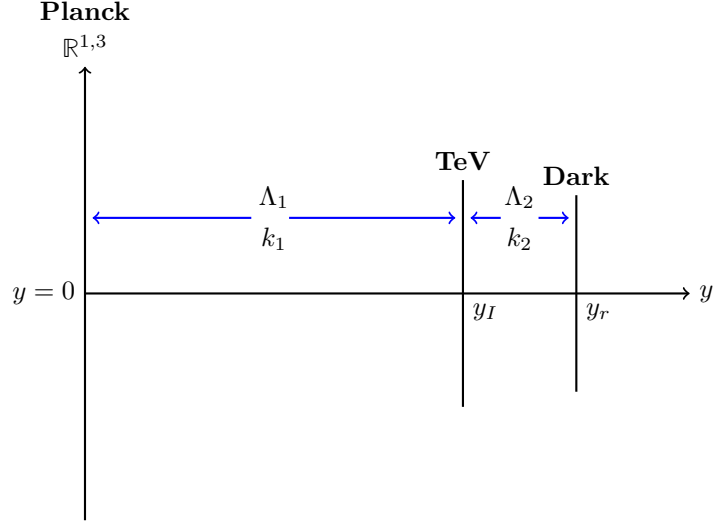

The remaining electroweak hierarchy problem of $\mu_\mathrm{UV} \sim$ TeV and $M_5 \sim M_\mathrm{Pl}$ can be addressed in a simple three brane extension to the RS model in the UV, illustrated in Fig. \ref{fig:UV}. For related work on multibrane extension to the RS solution, see \cite{Hatanaka_1999,Kogan_2002,Lee_2022,Girmohanta_2023}. We will again compactify the 5th dimension by the orbifold symmetry $S_1/\ZZ_2$ to the interval $y \in [0, y_r]$. In addition to the two branes located at the orbifold fixed points $y=0$, $y_r$, we add an intermediate brane at $y_I$. We identify the intermediate and IR branes with the UV and IR branes of the dark sector discussed previously. The brane at $y=0$, referred to as the Planck brane, has its energy scale set at $M_5$. In this parameterization, the gravitational action is
\begin{align}
    S = -\int \dd^4 x\ \dd y &\sqrt{|g|}\, \bigl\{2M_5^3R + \Lambda_1 [\theta(y)-\theta(y_I)] + \Lambda_2 [\theta(y_I)-\theta(y_r)] \bigr\} + \sqrt{|g_{\mathrm{ind}}|}\tau_0\delta(y) + \nonumber\\
    &\sqrt{|g_{\mathrm{ind}}|}\tau_1\delta(y-y_I) + S_{\mathrm{\Phi}} + \sqrt{|g_{\mathrm{ind}}|}\tau_2\delta(y-y_r) + S_{\mathrm{\Phi}}\,,
\end{align}
where $\Lambda_i$'s are the bulk cosmological constants, and $\tau_i$'s are the brane tensions. Using the ansatz $$\dd s^2 = e^{-2 A(y)}\tensor{\eta}{_\mu_\nu}\dd x^\mu \dd x^\nu - \dd y^2\,,$$
we find the warp factor 
\begin{equation}
A(y) =
    \begin{cases}
        k_1 y & 0 < y < y_I\\
        k_1 y_I + k_2 (y-y_I) & y_I < y < y_r
    \end{cases}\,.
\end{equation}
This is related to the Poincare coordinates used in previous derivations by $e^{A(y)}\dd y = \dd z$. The boundary conditions set $\Lambda_i = -24M_5^3 k_i^2$ in the bulk, and 
\begin{equation}
    \lambda_0 = 24M_5^3 k_1\,,\quad \lambda_1 = 12M_5^3(k_2-k_1)\,,\quad \lambda_2 = -24M_5^3 k_2\,,
\end{equation}
on the three branes respectively.

The 5D Planck mass is given by 
\begin{align}
    M_{Pl}^2 &= \frac{M_5^3}{k_1 k_2}\bigl(k_2 - (k_2-k_1)e^{-2k_1y_I} - k_1 e^{-2k_1 y_I - 2k_2(y_r-y_I)}\bigr)\nonumber\\
    &\approx \frac{M_5^3}{k_1}\,.
\end{align}

$k_1 y_I \approx 30$ generate the hierarchy between the Planck scale and the electroweak scale, and $k_2 (y_r-y_I)$ generate the hierarchy between the electroweak scale and $M\subtx{gap}$. For the Higgs portal COFI setup where $M_\mathrm{gap} \sim$ MeV, we require $k_2 (y_r-y_I) \approx 14$. In this three brane model, there are two radion degrees of freedom that acquire a VEV through some stabilizing mechanism. The Planck to intermediate brane region is stabilized by a radion with VEV $\sim$ TeV, and the intermediate to IR brane region, corresponding to the dark sector, is stabilized by the mechanism discussed previously with a VEV $\sim M_\mathrm{gap}$. The radion kinematics and the phase transition in a three brane cosmology are discussed in \cite{Lee_2022,Girmohanta_2023}, and is not the purpose of this work. We note that from the two radion kinetic terms, $k_2 > k_1$ is necessary to ensure the stability of the radions. We will additionally assume the radion stabilizing the Planck to electroweak hierarchy is heavy and thus can be integrated out in the description of the dark sector phase transition. This can be generically expected as the radion masses parametrically depend on the VEV.

\section{Phase Transition}
\label{sec:PT}

As the universe expands and cools, the energy of the gapped conformal dark sector eventually falls below the gap scale $M\subtx{gap}$, where the conformal symmetry is spontaneously broken and the dark sector undergoes a first-order confinement phase transition (PT). We study the dynamics of this PT in the weakly coupled holographic dual, where the cold, confined phase corresponds to the RS-like geometry established in Section \ref{sec:dual} where the free energy can be approximated by the dilaton effective potential of Eq. (\ref{eq:Veff}), 
\begin{equation}
    F_\mathrm{conf}(\chi) \approx V(\chi)\,.
\end{equation}

The deconfined phase in the early universe is dual to a black brane solution to Eq. (\ref{eq:EH}), approximately given by the AdS-Schwartzchild (AdS-S) metric, which is exact when taking $z_0 \rightarrow 0$, i.e. when the UV brane is taken to the conformal boundary. In this phase, the free energy as a function of the temperature $T$ is given by
\begin{equation}
    F_\mathrm{deconf}(T) = V_0 - 2\pi^4 (M_5^3/k^3) T^4\,.\label{eq:Fdecon}
\end{equation}

The PT proceeds via bubble nucleation, described by a bounce solution that interpolates between the local minima of the free energies of the confined and deconfined phases through a shared limit of $T \rightarrow 0$ and $\chi \rightarrow 0$, that is, removal of the IR boundary \cite{Creminelli_2002,Agashe_2020,Agashe_2021}. As the $d_-$ branch effective potential becomes singular in this common limit, we will limit the discussion of the phase transition to the branch of CFT with $d_+$ scaling. Despite the inevitable breakdown of 4D EFT at the shared limit, where the excited scalar modes become a continuum, the $d_+$ branch potential remains well defined so that we can still follow the analysis in \cite{Creminelli_2002}. Joining the free energies of the two phases at the shared limit, the constant of Eq. (\ref{eq:Fdecon}) is fixed to $V_0 = \frac{3N^2\lambda(2-\nu)}{2\nu} \braket{\chi}^4$. The critical temperature $T_c$ of the phase transition is set by $F_\mathrm{conf}(\braket{\chi}) = F_\mathrm{deconf}(T_c)$, which is
\begin{equation}
    T_c = \frac{\braket{\chi}}{\pi}\left(\frac{12\lambda(2-\nu)}{\nu}\right)^{1/4}\,.
\end{equation}
The phase transition can only complete when the rate of the phase transition $\Gamma \sim T^4 e^{-S_b}$, where $S_b$ is the Euclidean bounce action, surpasses the Hubble parameter, $H^4 \approx (\frac{\pi^2N^2 T_c^4}{24M_\mathrm{Pl}^2})^2$. The bounce action is thus bounded above by
\begin{equation}
    S_b \lesssim 4\left(\log{\frac{M_\mathrm{Pl}}{N T_c}}+\log{\frac{T}{T_c}}\right)\,.
\end{equation}
For Higgs portal PT, the gap scale and critical temperature of order MeV set the bound at $S_b \lesssim 190$.

We can analytically compute the $O(3)$ symmetric bounce action in the thin-wall regime, $T \sim T_c$, given by
\begin{align}
    S_b &= \frac{16\pi}{3}\frac{S_1^3}{(\Delta F)^2 T}\\
    &\approx \frac{N^2}{(3\lambda^3(1-T/T_c)^8)^{1/4}}\left[\left(\frac{\nu}{2-\nu}\right)^{3/4}\int_0^1\dd x\sqrt{\frac{1-x^{2\nu}}{\nu/2}-(1-x^4)}\right]^3\,. \label{eq:twbounce}
\end{align}
Compared to the numerically calculated bounce action\footnote{We thank Ameen Ismail for providing the code used for the numerical analysis.} \cite{Guada_2019,Guada_2020}, the error from the thin-wall approximation falls within the uncertainty estimated by varying the free energies in the regime where the bounce solution goes out of EFT control for supercooling of $T \sim 0.5 T_c$ \cite{Cs_ki_2023}. We will consider the COFI confinement PT in the minimal supercooling scenario where the PT promptly completes; therefore, it suffices to estimate the bounce action in the thin-wall limit. We study the PT for benchmark parameters $\tau_\mathrm{UV} = 3$, $\tau_\mathrm{IR} = -4\nu/100$. The small IR mass mistuning ensures the lightest KK mode is not excited for $\lambda \gtrsim 0.01$. We examine all viable parameters $\nu$, $M_\mathrm{gap}$ of COFI DM for the Higgs portal \cite{Hong_2020,Hong_2023} and choose $\lambda$ saturating the consistency constraints of Eq. (\ref{eq:lamspon}) and EFT validity, so that the bounce action $S_b \propto \lambda^{-3/4}$ is minimized.

\begin{figure}[t]
    \centering
    \includegraphics[width=.47\textwidth]{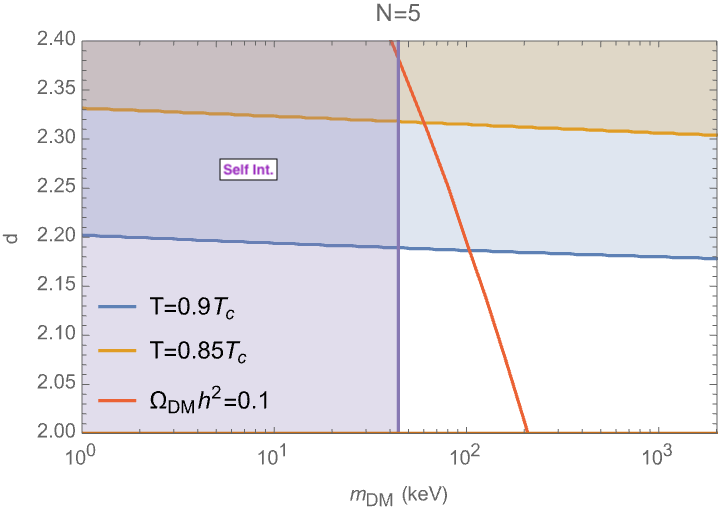}
    \qquad
    \includegraphics[width=.47\linewidth]{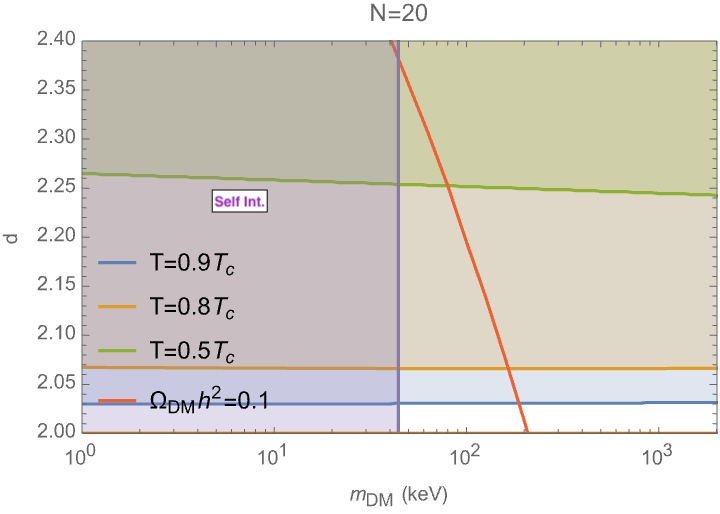}
    \caption{Theoretical/observational constraints for the Higgs portal COFI with a scalar mediator. PT completion bounds are shown for $N=5$ (left) and $N=20$ (right) for various degrees of supercooling. The red line indicates parameters where the correct dark matter relic abundance is obtained. Red line corresponds to the DM relic density matching the observed value. The region shaded in purple is excluded by the DM self-interaction bounds \cite{Hong_2020,Hong_2023}. The dark matter mass $m_\mathrm{DM} = 0.1 M_\mathrm{gap}$.}
    \label{fig:COFIHiggsPortal}
\end{figure}

\begin{figure}[h]
    \centering
    \includegraphics[width=.47\textwidth]{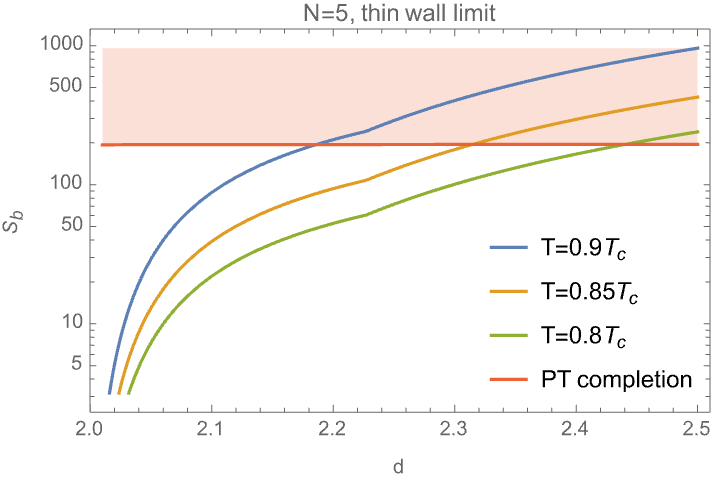}
    \qquad
    \includegraphics[width=.47\textwidth]{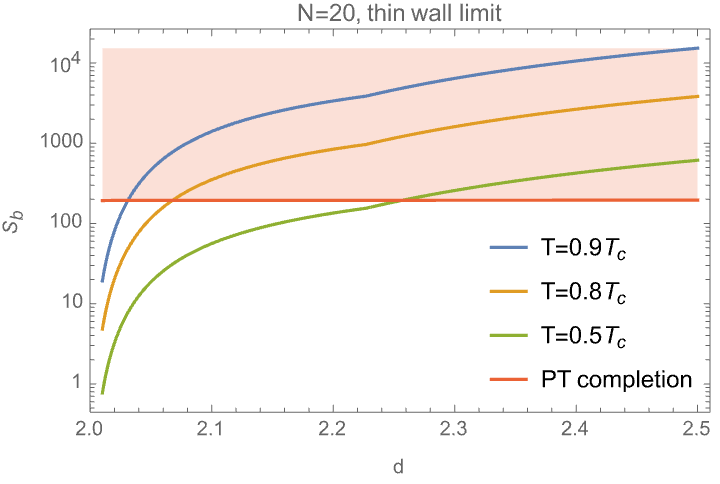}
    \caption{The bounce action estimated in the thin-wall limit for $d>2$, for parameters where the COFI mechanism produces corrrect relic density. In the shaded regions the phase transition does not complete.}
    \label{fig:ptsbounce}
\end{figure}

New constraints on the Higgs portal DM from the completion of PT are shown in Fig. \ref{fig:COFIHiggsPortal} for various degrees of supercooling and number of d.o.f $N=5, 20$ in the CFT. For a small number of d.o.f in the CFT, the PT completes for most of the allowed parameter space of Higgs portal with a scalar mediator\footnote{The viable parameter space for a vector mediator falls entirely under the $d_-$ branch CFT, so no PT bounds will be applied at this point.}, without requiring significant supercooling. This remains true when we consider a moderately large $N$, although a lower nucleation temperature will be necessary for PT completion. We show additionally, in Fig. \ref{fig:ptsbounce}, the bounce action evaluated at the viable Higgs portal parameters $d$, $M_\mathrm{gap}$ that produces the correct DM relic abundance. This is compared to the upper bound on the bounce action for PT completion. 

\begin{figure}[h]
    \centering
    \includegraphics[width=0.7\linewidth]{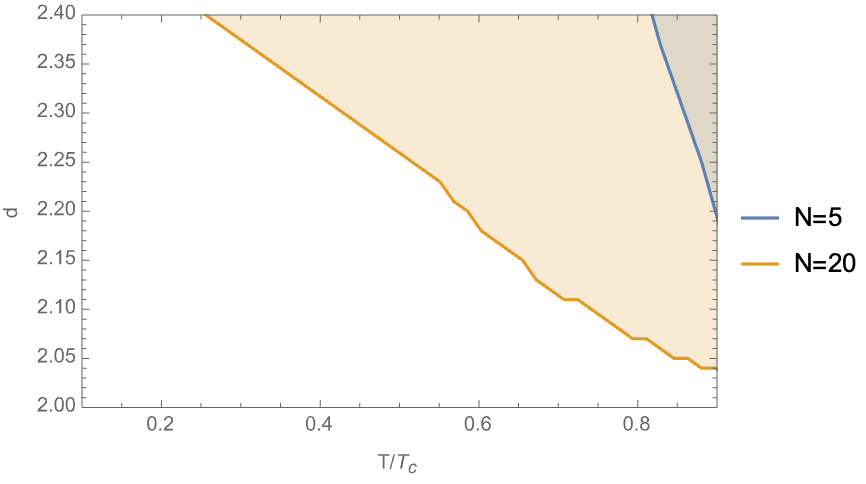}
    \caption{Exclusion plot for the minimum supercooling required for the completion of PT. In the shaded regions the phase transition does not complete.}
    \label{fig:supercool}
\end{figure}

Fig.~\ref{fig:supercool} shows the dependence on the amount of supercooling, $T/T_c$. We find that for a moderately large number of d.o.f, $N=20$, the Higgs portal parameter space receives no further constraints from PT completion for nucleation temperature $T \lesssim 0.3 T_c$.

Recently, first-order phase transitions, such as the one studied here, received a considerable amount of attention as a candidate source for the stochastic gravitational wave background observed by four independent pulsar timing arrays (PTA)~\cite{NANOGrav:2023gor,Xu:2023wog,Reardon:2023gzh}. The fit to the observed gravitational wave spectrum prefers reheat temperatures of order 20~MeV$-$3~GeV~\cite{NANOGrav:2023hvm}. The phase transition in the simplest COFI model with a Higgs portal occurs at a temperature of order $M_{\rm gap}\sim$ a few MeV, intriguingly close to the preferred range. However, in the COFI model, the dark sector is colder than the SM, $T_d/T_{\rm SM}\sim 0.1$~\cite{Hong_2020}. The density of stochastic gravitational waves is proportional to the square of the parameter $\alpha$, the ratio of the latent heat released in the phase transition to the energy of the surrounding radiation bath~\cite{Caprini:2015zlo,Caprini:2019egz}. Without significant supercooling, $\alpha\sim T_d^4/T_{\rm SM}^4\sim 10^{-4}$, making the signal too feeble to be observed. Thus, very large supercooling, even larger than in the models where the dark sector and the SM are in thermal equilibrium\footnote{Many such models have been discussed in the literature. For an example of a construction that shares many of the features of our setup, see Ref.~\cite{Ferrante:2023bcz}.}, would be required to fit the PTA data. This does not generically occur in the model we studied here.

\section{Conclusions}
\label{sec:conc}

Conformal field theories are generic on the landscape of quantum field theories, occurring whenever renormalization group evolution drives a theory towards an interacting fixed point. It is therefore well-motivated to consider the possibility that Beyond the Standard Model physics takes the form of a dark sector described by a CFT in a broad energy range. Conformal Freeze-In (COFI) scenario shows that a dark matter candidate particle can arise naturally in this scenario as a bound state of the CFT degrees of freedom, confined in the IR thanks to a ``portal" interaction mediating conformal symmetry breaking in the SM sector to the dark sector. The relic density of this particle can be calculated in terms of a small number of parameters, and can match the observed DM abundance for appropriate model parameters (in particular, $m_{\rm DM}$ in the phenomenologically interesting keV-MeV range). 

Previous studies of the COFI scenario used a 4D description, where many calculations are inherently difficult due to strong coupling in the CFT. AdS/CFT duality suggests that an alternative, dual description of the same physics can be provided by a weakly-coupled theory with an additional spatial dimension. In this paper, we proposed an explicit 5D theory which provides a dual of the COFI scenario. Our construction is summarized in Fig.~\ref{fig:5D}. Conformal symmetry in 4D maps into the AdS geometry of the 5D space. The scale of the conformal symmetry breaking in the SM (in the case of the Higgs portal, the weak scale) is identified as the location of the UV brane. The scale of confinement in the dark sector (the ``gap scale") is the location of the IR brane. The renormalization group flow of the CFT operator sourced by the Higgs VEV and inducing confinement is described by the dynamics of a bulk scalar field (the radion) coupled to the Higgs field on the UV brane, with a mass term on the IR brane. The hierarchy between the gap and electroweak scale is dual to the relative location of the two branes, determined by minimizing the radion potential. 

As an example of a calculation that can be carried out in the 5D setup, we considered the dynamics of the cosmological phase transition in the dark sector between the high-temperature conformal and low-temperature confined phases. Once the 5D dual has been constructed, this calculation becomes closely parallel to the previous analyses of phase transitions in Randall-Sundrum model and related setups. We observe that our model typically undergoes the confining phase transition without significant supercooling, avoiding the risk of slipping into an eternally-inflating regime inconsistent with standard cosmology.  

Depending on the details of the CFT, confinement in the dark sector can result in a variety of different theories below the gap scale. The COFI scenario made some basic assumptions about physics below the gap scale, but is rather agnostic to details of this physics. In the 5D dual, the physics around and below the gap scale is modeled by the IR brane. In this paper, we considered the simplest possibility of a point-like IR brane, with the DM as a brane-localized field. Many other scenarios are possible, including for example non-point-like ``soft walls" \cite{Karch:2006pv, Batell:2008zm, Falkowski:2008fz,Cabrer:2009we, Megias:2018sxv}. These may correspond to more complex DM sectors, or perhaps even gapped continuum states realizing the framework of continuum DM~\cite{Csaki:2021gfm,Csaki:2021xpy,Fichet:2022ixi,Fichet:2022xol}. We leave the study of these possibilities for future work.

%%%%%%%%%%%%%%%%%%%%%%%%%%%%%%%%%%%%%%%%%%%%%%%%%%% 
\section*{Acknowledgements}

We would like to thank Csaba Csaki, Steven Ferrante, Thomas Hartman, Zamir Heller-Algazi, Sungwoo Hong, Jay Hubisz, and Ameen Ismail for useful discussions. Our research is supported by the NSF grant PHY-2309456. LL is also supported by the NSERC Fellowship PGS D - 578021 - 2023.

%%%%%%%%%%%%%%%%%%%%%%%%%%%%%%%%%%%%%%%%%%%%%%%%%%% 

\input{dEFT}

\bibliographystyle{utphys}
\bibliography{cite}

\end{document}

%% file: dEFT.tex
\appendix

\section{Dilaton kinetic term} \label{ap:kin}
\label{sec:app}

We compute the kinetic term for the dilaton effective action from the 5D perspective by decomposing the 5D Ricci scalar using a metric ansatz involving a scalar degree of freedom to 2 derivative order \cite{GOLDBERGER2000275,csaki2004tasi,Cs_ki_2022,Bellazzini_2014}. In the most general form, we can parameterize the fluctuations of the metric as 
\begin{equation}
    \dd s^2 = \frac{1}{[A(z)(1+F(x, z))]^2}\left((\eta_{\mu\nu} + h_{\mu\nu}(x))\dd x^\mu \dd x^\nu - (1+G(x, z))^2\dd z^2\right)\,.
\end{equation}
To linear order in the scalar fluctuation $F(x,z)$, the 4D off-diagonal components of the Einstein equations are satisfied if $G = 3F$. The tensor fluctuation is parametrically $h \sim (\partial F)^2$, so its contribution to the action appear at $\mathcal{O}(\partial^4)$ and can be omitted at the order of the kinetic term \cite{Cs_ki_2022}. We will also assume negligible backreaction, that is, $A(z)=kz$. The scalar $F(x, z) = r(x)g(z)$ then satisfies the equation of motion
\begin{equation}
    z g'(z) - 2g(z) = 0\,.
\end{equation}
The solution gives the wave function of the radion $r(x)$ in the absence of stabilizing mechanism,
\begin{equation}
    g(z) = N z^2\,,
\end{equation}
where $N$ is a normalization constant. The metric ansatz then takes a simpler form,
\begin{equation}
    \dd s^2 = \frac{1}{[A(z)(1+Nz^2r(x)]^2}\left(\eta_{\mu\nu}\dd x^\mu \dd x^\nu - (1+3Nz^2r(x))^2\dd z^2\right)\,.
\end{equation}

The 5D gravitational action decomposes as
\begin{align}
    S_\mathrm{grav} &= -\int \dd^5 x\sqrt{g}\ (2M_5^3R + \Lambda)\\
    &\supset -2M_5^3 \int \dd^4 x\int \dd z\ \frac{(1+3F)}{[kz(1+F)]^5}[14k^2z^2F\partial^2F + 4k^2z^2(\partial F)^2 + 2k^2z^2 \partial^2F]\\
    &= 2M_5^3 \int \dd^4 x\int \dd z\ \frac{1}{(kz)^5} 6k^2z^2 N^2z^4 (\partial r)^2 + \mathcal{O}(\partial^3)\,.
\end{align}

We normalize the radion as $F(x, z) = \frac{z^2}{z_1^2}r(x)$ and obtain an effective 4D Lagrangian by integrating over the orbifold,
\begin{align}
    \mathcal{L}_{r} &= \frac{24M_5^3}{k^3} \int_{z_0}^{z_1}\dd z\ \frac{z}{z_1^4} (\partial r)^2\\
    &\approx \frac{12M_5^3}{k^3}(\frac{1}{z_1}\partial_\mu r(x))^2\,,
\end{align}
where $\braket{\chi} = \frac{1}{z_1}$ is identified as the dilaton VEV and $r(x)$ is the 4D dilaton field. The action is canonically normalized by taking $\Tilde{\chi} \equiv \sqrt{24M_5^3/k^3}\frac{1}{z_1}r(x)$.